\documentclass[twocolumn]{aastex701}

\usepackage{amsmath}
\usepackage{graphicx}
\usepackage{booktabs}
\usepackage{caption}
\usepackage{comment}
\usepackage{multirow}
\usepackage[absolute,overlay]{textpos}

\usepackage[dvipsnames]{xcolor} 
\usepackage{longtable}

\newcommand{\hi}{\texttt{M2e4\_mu0.4}}
\newcommand{\med}{\texttt{M2e4}}
\newcommand{\lo}{\texttt{M2e4\_mu4.2}}
\newcommand{\msun}{M$_\odot$}

\shorttitle{Star-Forming Gas in STARFORGE}
\shortauthors{Kaalva et al.}

\begin{document}

\title{The Evolution of Star-Forming Gas in STARFORGE: From Clouds, to Cores, to Stars}

\author[0000-0000-0000-0000]{Ananya Kaalva}
\affiliation{Department of Astronomy, The University of Texas at Austin, Austin, TX 78712, USA}
\email{offner@gmail.com}
\author[0000-0003-1252-9916]{Stella S. R. Offner}
\affiliation{Department of Astronomy, The University of Texas at Austin, Austin, TX 78712, USA}
\email{soffner@astro.as.utexas.edu}
\author{Nina Filippova}
\affiliation{Department of Astronomy, The University of Texas at Austin, Austin, TX 78712, USA}
\email{soffner@astro.as.utexas.edu}
\author{Michael Y. Grudi\'c}
\affiliation{Center for Computational Astrophysics, Flatiron Institute, 162 Fifth Avenue, New York, NY 10010, USA}
\email{soffner@astro.as.utexas.edu}

\begin{abstract}
Star formation occurs within dense regions of giant molecular clouds (GMCs), however, exactly how gas collects and evolves to form individual stars and what role dense cores play 
remains unclear. 
We use the Lagrangian cell information in the STARFORGE simulation suite to track star-forming gas in three GMCs with varying magnetic field strengths. We find that, once a protostar forms, the lifetime of the unaccreted gas  
 correlates with the final stellar mass, where low-mass stars ($M_{*}<0.5M_{\odot}$) 
accrete for 0.5-0.6~Myr from a relatively local reservoir of gas, and  high-mass stars ($M_{*}>2M_{\odot}$) accrete over 
3.3-4.7~Myr from a much larger volume. Although the protostellar accretion time increases weakly with magnetic field strength, the accreting gas radii, velocity dispersions, virial parameters, and magnetic energy ratios are largely insensitive to the global cloud properties. 
At the time of protostar formation, the unaccreted gas exhibits linewidth-size and mass-size relations characteristic of turbulently regulated, isothermal dense cores, following $\sigma_v \propto R^{0.47-0.55}$ and $M \propto R^{1.0-1.1}$, respectively. Low- and intermediate-mass stars undergo relatively continuous accretion and their accretion histories are well-fit by either isothermal sphere, turbulent core, or competitive accretion models, where no one model fits all masses. However, many high-mass stars experience intermittent accretion and their accretion histories are not well-fit by any of these models. 
While the distribution of accreting gas is more extended than typically-defined dense cores, the physical properties and structure of the star-forming gas resemble those of observed cores and are largely regulated by turbulence and feedback. 
\end{abstract}


\section{Introduction}

The general picture of star formation is well established: stars form through the gravitational collapse of dense regions within giant molecular clouds (GMCs), which fragment into filaments and cores that eventually produce stellar systems \citep{PinedaArzoumanian2023a}. However, the collapse of GMCs is itself highly complex, shaped both by global cloud dynamics and local variations in density, velocity, and magnetic structure often produced by the stars themselves. Star formation within these clouds is further complicated by the non-linear interplay between self-gravity, supersonic turbulence, magnetic fields, and stellar feedback \citep{girichidis_physical_process}. These processes act together to regulate when and where gas collapses, thus influencing the star formation rate, initial mass function (IMF), and structure of emerging star clusters \citep{OffnerClark2014a,Federrath2015, Mandal2021,GuszejnovMarkey2022a,FariasOffner2024a}. 

Observationally, the low star formation efficiency of GMCs indicates the importance of regulatory mechanisms, including stellar feedback in the form of protostellar outflows, radiation, winds, and supernovae, which appear to play a dominant role in halting star formation and dispersing the cloud \citep{GrudicGuszejnov2022a,Guszejnov2022}. Although the role of feedback on large scales within GMCs is recognized, its impact on the earliest phases of the collapse and evolution of individual star-forming cores remains less clear \citep{NeralwarColombo2024a}. Simulations suggest that protostellar outflows drive local turbulence and expel dense gas from the natal core, thereby reducing stellar masses and shaping the IMF \citep{Federrath2015,OffnerChaban2017a,guszejnov_starforge_2021}. This feedback also shapes the properties and evolution of dense cores, both starless and protostellar \citep{NeralwarColombo2024a,OffnerTaylor2025a}.

However, dense cores represent only a partial view of the gas  that collapses and ultimately forms stars, as a significant fraction of the identified over-densities is either dispersed or never collapses to form a star \citep{OffnerChaban2017a,OffnerTaylor2022a}.
Gaining a more complete picture of how feedback, turbulence, and gravity interact in the star formation process requires following the evolution of the true star-forming gas over time. Some numerical studies have studied this gas by identifying gas over-densities and tracking their properties \citep[e.g.,][]{SmithClark2009a, GongOstriker2015a,SmullenKratter2020a,OffnerTaylor2022a,OffnerTaylor2025a}. However, it is difficult to disentangle gas that is accreted by a star -- truly star-forming gas -- and gas that is part of the star-forming ``core" that is eventually dispersed.
A few prior studies have employed Lagrangian tracer particles, which follow the trajectories of individual gas parcels \citep{MoczBurkhart2018a,Pelkonen2021,CollinsLe2023a,CollinsLe2024a}, however, none of these included stellar feedback mechanisms.

The STARFORGE simulation suite addresses this gap by tracking gas in a Lagrangian fashion and by including all fundamental physical processes relevant to star formation in GMCs, including magnetohydrodynamics (MHD), gravity, radiative transfer, and all major forms of stellar feedback. \citep{Grudic2021}.\footnote{http://www.starforge.space/} \citet{OffnerTaylor2025a} used the gas density to identify and follow dense cores over time in the STARFORGE simulations, demonstrating how feedback regulates core properties and their dispersal. Here, we make use of the Lagrangian information to track the gas that goes on to form individual protostars. 

We evaluate how key properties such as mass, velocity dispersion, radius, and energy vary with respect to time.  This approach allows us to directly address several open questions: 
How do the properties of star-forming gas evolve as collapse proceeds, and to what extent is this evolution shaped by stellar feedback? How does low-mass and high-mass star formation differ? How do the properties of the star-forming gas compare to those of identified dense cores?
We describe the simulations and methods used to trace the star-forming gas in Section \ref{methods}. We present our analysis of the lifetimes, mass functions, properties, and accretion in Section \ref{results}. We discuss the results in the context of prior work in Section \ref{discussion} and summarize in Section \ref{conclusion}.

\section{Methods} \label{methods}

\subsection{STARFORGE Numerical Simulations}\label{sec:starforge}

We analyze three magnetohydrodynamic (MHD) simulations with different magnetic field strengths from the STARFORGE project \citep{Grudic2021,Guszejnov2022}. Each models the collapse and star formation activity in a $2\times 10^4$\,\msun~GMC and uses identical initial conditions aside from the strength of the initial magnetic field, which varies by an order of magnitude between the three runs. The simulations are performed using the  GIZMO magnetohydrodynamics code,\footnote{http://www.tapir.caltech.edu/ phopkins/Site/GIZMO.html} which employs a Lagrangian meshless finite-mass method. Stellar feedback processes, including protostellar outflows, stellar radiation, stellar winds, and core-collapse
supernovae, are modeled self-consistently and coupled to
the evolution of individual stars, which are represented
by sink particles that accrete gas over time. The simulations evolve until stellar feedback disperses the cloud,
marking a natural termination of star formation. Full descriptions of the numerical framework and the physics modules are provided in \citet{Grudic2021}. In this study, we focus on the evolution of gas that forms individual stars, analyzing the prestellar and protostellar phases of accreting material across three runs with different magnetic field strengths. Here we give a brief overview of the simulation properties and refer the reader to \citet{Guszejnov2022} for a more detailed discussion of the simulations.

Each simulation is initialized with a uniform density spherical GMC of mass $M_0 = 2\times10^4\text{ }M_\odot$, radius $R_0 = 10$ pc, and a uniform magnetic field along the $z$-axis. The gas is discretized with a mass resolution of $\Delta m = 10^{-3}$\,\msun. The initial gas and dust temperatures are set by the solar neighborhood interstellar radiation field \citep{Mathis1983}. The cloud is embedded in an ambient medium that is 100 times lower density than the cloud to approximate thermal pressure equilibrium. We follow the STARFORGE naming convention for the three simulations, where \texttt{M2e4} (fiducial), \texttt{M2e4\_mu4.2}, and \texttt{M2e4\_mu0.4} correspond to clouds with initial magnetic field strengths of $B_z$=6.3, 2, and 20 $\mu$G, respectively.

Turbulent motions are initialized with a random velocity field following a power spectrum $E_k \propto k^{-2}$ and normalized to achieve a virial parameter $\alpha_{\rm vir} = 2$, yielding a three-dimensional velocity dispersion $\sigma_{v} = 3.2$ km/s. Star formation is modeled via sink particles that form when gas exceeds a density threshold of $n_{\rm H} \sim 10^{10}$ cm$^{-3}$ and satisfies the local Jeans and tidal collapse criteria. Table \ref{tab:simparams} summarizes the simulation properties and final run times.

\renewcommand{\arraystretch}{1.3}

\begin{table*}
\centering
\begin{tabular}{lccccc}
\hline
\textbf{Cloud Label} & $\mu$ & $\beta$ & $E_B/|E_{\text{grav}
}|$ & $B_z\text{ }(\mu\text{G})$ & Cloud Lifetime (Myr)\\
\hline
\lo & 4.2 & 0.78 & 0.01 & 2 & 11.2\\
\med (fiducial) & 1.3 & 0.078 & 0.1 & 6.3 & 12.1 \\
\hi & 0.42 & 0.0078 & 1 & 20 & 17.9\\
\hline
\end{tabular}
\caption{Initial conditions for the three STARFORGE simulations analyzed, where $\mu$ is the mass-to-magnetic flux ratio, the plasma $\beta$ parameter is the ratio of the thermal to magnetic pressure $\beta= P_{\rm thermal}/P_{\rm magnetic}$ given an initial 10~K gas temperature, and $B_z$ is the initial magnetic field strength.  
The final column lists the cloud lifetime, at which point star formation has ceased. All runs begin with total cloud mass $M_0 = 2 \times 10^4\, M_\odot$ and radius $R_0 = 10$ pc, and have mass resolution $\Delta m = 10^{-3}\, M_\odot$. }
\label{tab:simparams}
\end{table*}

\subsection{Tracking Star-Forming Gas}

To study the evolution of gas involved in individual forming stars, we use the cell IDs to isolate and track the subset of gas cells that accrete onto each sink particle. We exclude pure feedback cells, i.e., recently launched outflow and wind material, from the calculation of the properties; these contribute little mass but tend to have high velocities. To remove these, we apply a mass filter that excludes cells with masses below the initial gas resolution of $10^{-3}$~\msun. 
We then extract the positions, velocities, densities, and magnetic properties of the accreted gas across all snapshots. We define the prestellar phase as the interval prior to the formation of the protostar, during which the total mass of the subset remains constant. We define the protostellar phase as the period of active accretion, which continues as long as the remaining tracked gas within the simulation domain exceeds $0.01$~\msun. Below this threshold, some stars experience a small amount of residual accretion at late times in the simulation.  We exclude these cells, since they introduce stochasticity in the calculated properties due to small-number statistics.

\subsection{Gas Properties}\label{sec:properties}

For each gas subset that forms each star, we compute six properties as functions of time, including the gas mass, $M$, mass-weighted velocity dispersion, $\sigma_{\rm v}$, and effective radius, $r_{\rm eff}={\sqrt[3]{{3(\sum_{i} {m_{i}/\rho_{i}})}/{4\pi}}}$, where $\rho_{i}$ is the density of each cell. This is the radius of a sphere with volume equivalent to the structure volume.

We also compute the virial parameter, $\alpha_{\rm vir}=2E_{\rm KE}/|E_{\rm PE}|$, given by twice the ratio of the kinetic energy to the gravitational potential energy of the subset. The kinetic energy is calculated as $E_{\rm KE} = \sum_{i} {{1\over{2}} m_{i}(\mathbf{v}_{i}-\overline{v})^{2}}$, where $\bar v$ is the center-of-mass velocity. The gravitational potential energy is given by $E_{\rm PE} = {1\over{2}}\sum_{i}m_{i}{\phi_{i}}$, where $\phi_{i}$ is the gravitational potential of each cell. We calculate this property using the {\it pytreegrav} package \citep{Grudic2021_pytreegrav}. Finally, the magnetic energy is calculated using $E_{\rm BE} = \frac{1}{8\pi}\sum_{i}{{{\mathbf B}_{i}}^{2}}m_{i}/{\rho_{i}}$, where ${\mathbf B}_i$ is the magnetic field value for each cell. 

\section{Results}

\label{results}

\subsection{Star Formation Lifetimes}

\begin{table*}
\centering
\begin{tabular}{llcccc}
\hline
\textbf{Phase} & \textbf{Cloud Label} & \textbf{Overall (Myr)} & \textbf{Low Mass (Myr)} & \textbf{Intermediate Mass (Myr)} & \textbf{High Mass (Myr)} \\
\hline
\multirow{3}{*}{Prestellar} 
& \lo & $4.50^{+1.46}_{-1.11}$ & $4.11^{+1.30}_{-0.95}$ & $5.14^{+1.49}_{-1.12}$ & $5.19^{+1.12}_{-0.94}$ \\
& \med & $5.71^{+1.68}_{-1.31}$ & $5.20^{+1.85}_{-1.05}$ & $6.42^{+1.33}_{-1.28}$ & $6.31^{+1.38}_{-1.23}$ \\
& \hi & $9.55^{+2.82}_{-2.58}$ & $9.24^{+3.08}_{-2.47}$ & $10.12^{+2.35}_{-2.45}$ & $9.68^{+2.63}_{-2.51}$ \\
\hline
\multirow{3}{*}{Protostellar} 
& \lo        & $0.20^{+0.35}_{-0.12}$ & $0.15^{+0.12}_{-0.10}$ & $0.37^{+0.44}_{-0.20}$ & $0.77^{+0.94}_{-0.37}$ \\
& \med & $0.20^{+0.30}_{-0.10}$ & $0.15^{+0.12}_{-0.11}$ & $0.40^{+0.42}_{-0.20}$ & $0.88^{+0.83}_{-0.36}$ \\
& \hi & $0.25^{+0.35}_{-0.12}$ & $0.17^{+0.15}_{-0.07}$ & $0.64^{+0.66}_{-0.30}$ & $1.74^{+1.02}_{-0.70}$ \\
\hline
\end{tabular}
\caption{Median lifetimes with interquartile range  for the gas subsets separated by star-formation phase and stellar mass, where we define low-mass stars as those with $M \leq 0.5$ \msun, intermediate stars as those with  0.5 \msun$< M < 2.0$ \msun, and high-mass stars as those with $M \geq 2$ \msun. Within each simulation, lifetimes are reported as both the median of the overall cloud population and binned by the final stellar masses.}
\label{tab:lifetimes}
\end{table*}

\begin{figure*}
    \centering
    \includegraphics[width=1\linewidth]{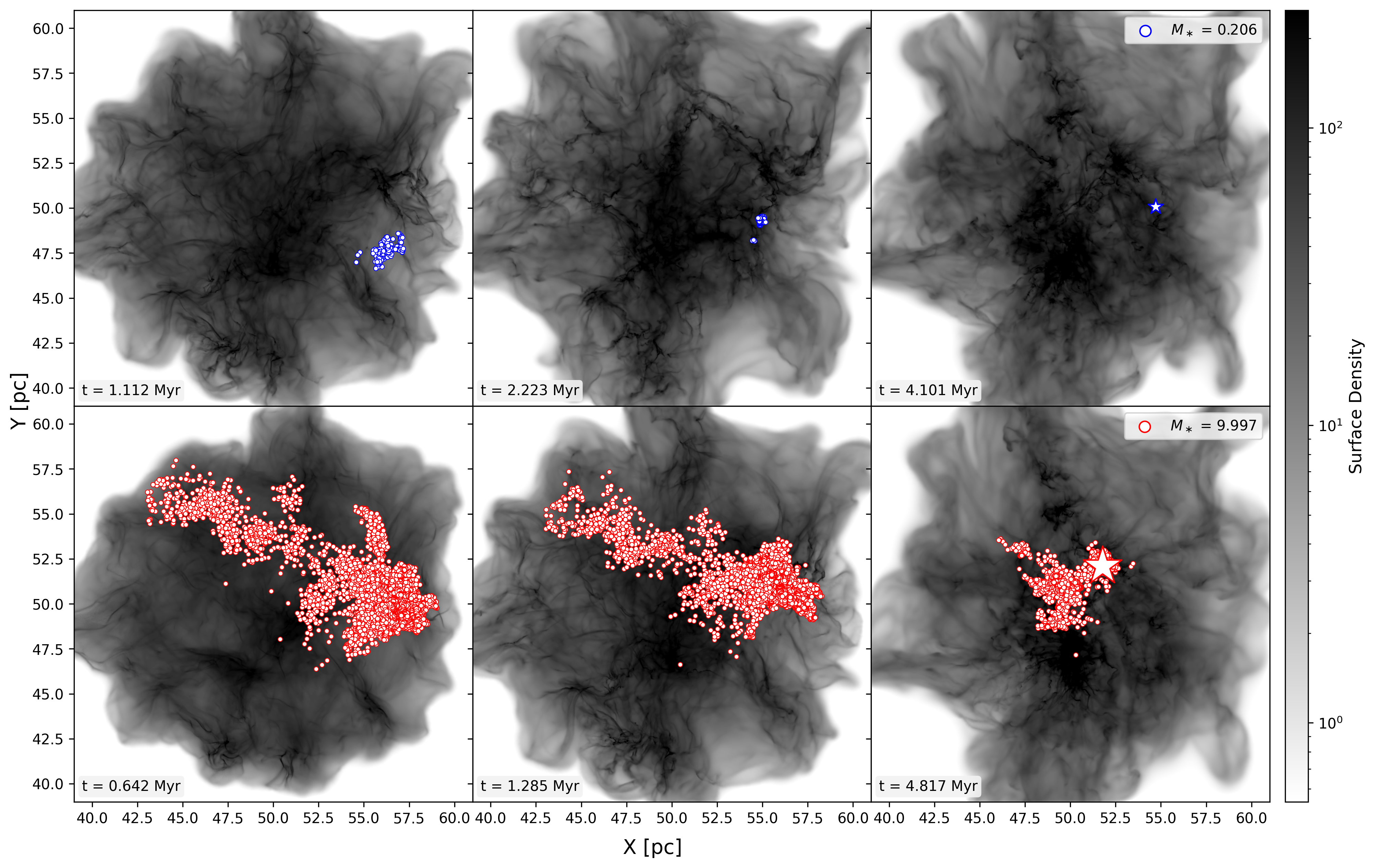}
    \caption{ 
    Gas forming typical individual low-mass and high-mass stars during the prestellar phase at three times in an \med\ cloud. The top row represents the spatial distribution of gas  forming a low-mass star with final mass $0.206$\,\msun\ and the bottom row is the gas for a high-mass star of $9.997$\,\msun. Gray areas indicate the spatial density of the cloud gas, colored circles represent the star-forming gas, and the star symbols indicate the location forming star, sized proportionally to its mass.
\label{fig:ex_reservoir_pre}}
\end{figure*}

\begin{figure*}
    \centering
    \includegraphics[width=1\linewidth]{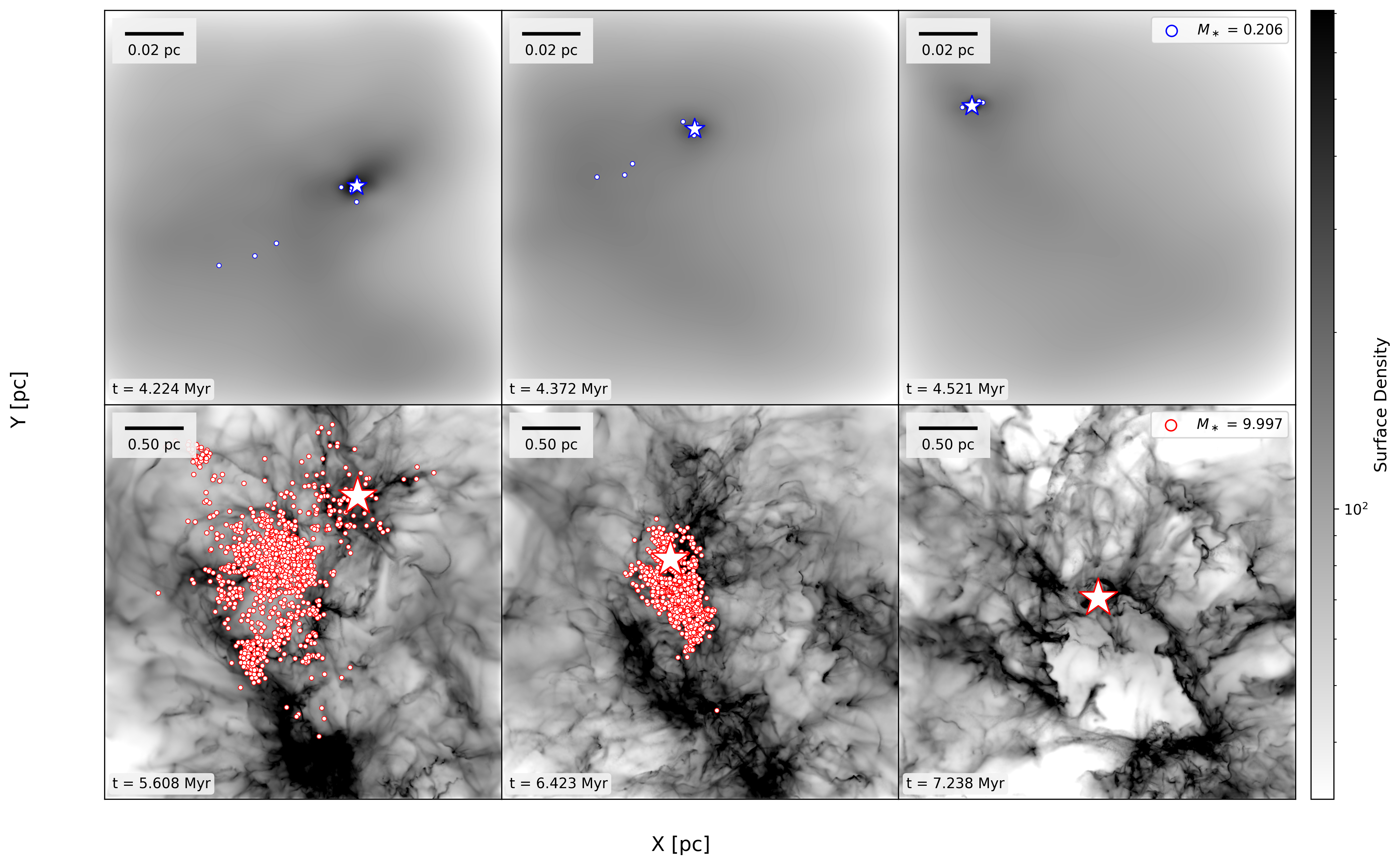}
    \caption{Gas forming typical individual low-mass and high-mass stars during the prestellar phase at three times in an \med\ cloud. The top row represents the spatial distribution of gas  forming a low-mass star with final mass $0.206$\,\msun\ and the bottom row is the gas for a high-mass star of $9.997$\,\msun. Gray areas indicate the spatial density of the cloud gas, colored circles represent the star-forming gas, and the star symbols indicate the location forming star, sized according to its mass. 
    \label{fig:ex_reservoir_post}}
\end{figure*}

To characterize the timescales over which stars form, we measure the average duration for which the gas remains on the domain in both the prestellar and protostellar phases. Table \ref{tab:lifetimes} shows these timescales for each simulation, where we also subdivide the sample by the initial mass of the gas subsets, i.e., the masses of the final stellar population. We define low-mass, intermediate-mass, and high-mass stars as $M \leq 0.5$\,\msun,  $0.5$\msun\,$<M<2$\msun, and $M \geq 2$\,\msun, respectively.

We find that the median prestellar lifetime is approximately 4.5, 5.7, and 9.6 Myr for the three magnetic field strength calculations, respectively. These lifetimes increase systematically with the initial magnetic field strength, since star formation begins later and continues longer in the stronger field runs \citep[see][]{GuszejnovGrudic2022a}. Meanwhile,  there is a weak relationship between stellar mass and the prestellar lifetime,  
indicating that high-mass stars take longer to form in each calculation. Together, these trends suggest that the prestellar phase is largely determined by the global cloud environment.

Table \ref{tab:lifetimes} shows that the median protostellar lifetime is approximately 0.20 Myr, where the protostellar lifetime in the strong-field case, 0.25 Myr, is only slightly longer than in the two weaker field runs. 
However, these lifetimes strongly scale with stellar mass, meaning that high-mass stars accrete over significantly longer timescales in all three clouds. Low-mass stars form in $\sim$0.15 Myr, and intermediate-mass stars in $\sim 0.4-0.6$ Myr. High-mass stars accrete $\sim 0.8-1.8$ Myr on average, approximately five times longer than their low-mass counterparts. This is due to the larger quantity and more extended spatial distribution of the accreting gas cells. Figure \ref{fig:ex_reservoir_pre} shows an example of the differences in gas reservoirs between a low- and high-mass star during early times in the simulation, and Figure \ref{fig:ex_reservoir_post} shows the differences in accretion between the two stars during the protostellar phase. Comparing the timescales of the different panels indicates that the gas accreting onto the higher-mass star is both initially more extended and remains extended for a longer period of time. Meanwhile, the low-mass star accretes from a relatively local gas reservoir. 

The size of the gas distributions imposes a characteristic timescale,  $t_{\rm acc} \sim \ell / \sigma_{v}$, to accrete gas from distance, $\ell$, where $\sigma_v = 3.2 (\ell / {\rm 10~pc})^{0.5}$ km/s given a global cloud velocity dispersion of 3.2\,km\,s$^{-1}$ (see \S\ref{sec:starforge}). A 0.1\,pc size structure and a 2\,pc structure will have accretion crossing times of 0.3\,Myr and 1.4\,Myr, respectively. These times are comparable to the median accretion times in Table \ref{tab:lifetimes} for forming low- and high-mass stars.
Examples of additional protostellar gas distributions are shown in Appendix 
\ref{append:gas_dist}.

\subsection{Core Mass Function}

\begin{figure*}
    \centering
    \includegraphics[width=1\linewidth]{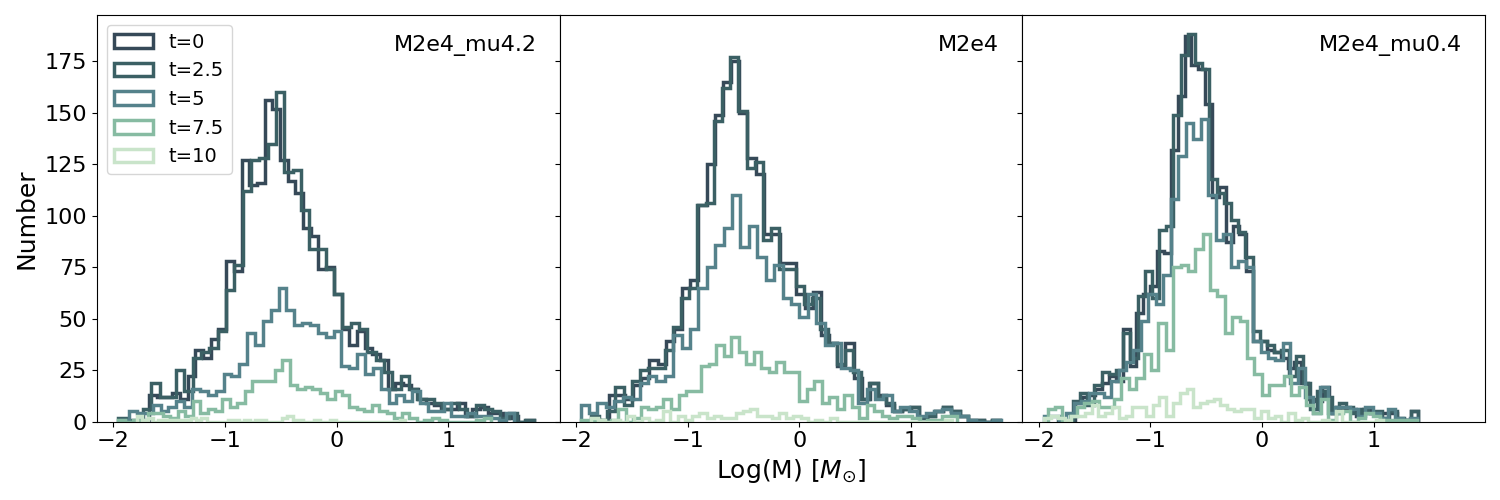}
    \caption{Time evolution of the prestellar gas mass function (PGMF) across the three STARFORGE simulations with varying magnetic field strengths (\lo, \med, \hi). Each panel shows the log-binned distributions of gas masses at five time snapshots, with line shading indicating time. Counts represent the number of prestellar gas subsets present per mass bin at each snapshot.}
    \label{fig:coremass}
\end{figure*}

We examine the time-evolution of the prestellar gas mass function (PGMF) across all three simulations in Figure \ref{fig:coremass}, which captures the changing population of star-forming gas over the course of the GMC evolution. Here, the masses represent any non-accreted gas, so the distribution includes both prestellar and protostellar gas subsets. 

These distributions, while exhibiting a similar log-normal shape, are distinct from canonical core mass functions \citep[e.g.,][]{AlvesLombardi2007a,EnochEvans2008b,KonyvesAndre2015a,BettiGutermuth2021a}. The prestellar gas associated with low-mass stars is often strongly concentrated around the forming star, and the morphology resembles an observationally defined core (see Fig.~\ref{fig:ex_reservoir_pre}). However, the prestellar gas distributions of high-mass stars are generally very extended, and any apparent compact core around these forming stars encompasses only a small amount of the actual gas that will be accreted. Conversely, not all gas within the parent over-density is ultimately accreted by the embedded forming star, since some of the material is entrained and ejected by stellar feedback or accreted by a nearby stellar companion. 

At early times ($t \leq 2.5$ Myr), the PGMF in all three clouds resembles a log-normal with a tail to higher masses and is sharply peaked around the median stellar mass $M\approx 0.3$ \msun. Since few stars have begun forming at this point, the PGMF reflects the stellar IMF. As the cloud evolves and stars form, the number of remaining gas subsets declines, and the PGMF becomes progressively less populated by intermediate- and low-mass stars.  Since these lower mass stars have shorter accretion times, they rapidly accrete their gas and are no longer represented in the PGMF. In contrast, the shape of the PGMF at higher masses ($M \gtrsim 3$ \msun) remains very similar from 0-7.5 Myr. Lower-mass subsets remaining on the domain represent either long-accreting stars that are nearing their final mass or late-start forming stars, and the PGMF flattens. 

In the \hi\ cloud the stronger magnetic support delays collapse and suppresses fragmentation, so the PGMF evolves more slowly. 
The initial shape of the PGMF is qualitatively similar across all three clouds, as expected for a largely invariant IMF \citep{GuszejnovGrudic2022a}, and we find that the cloud magnetic field strength mainly influences the duration of the protostellar phase.

\subsection{Gas Properties}

\begin{figure}
    \centering
    \includegraphics[width=1\linewidth]{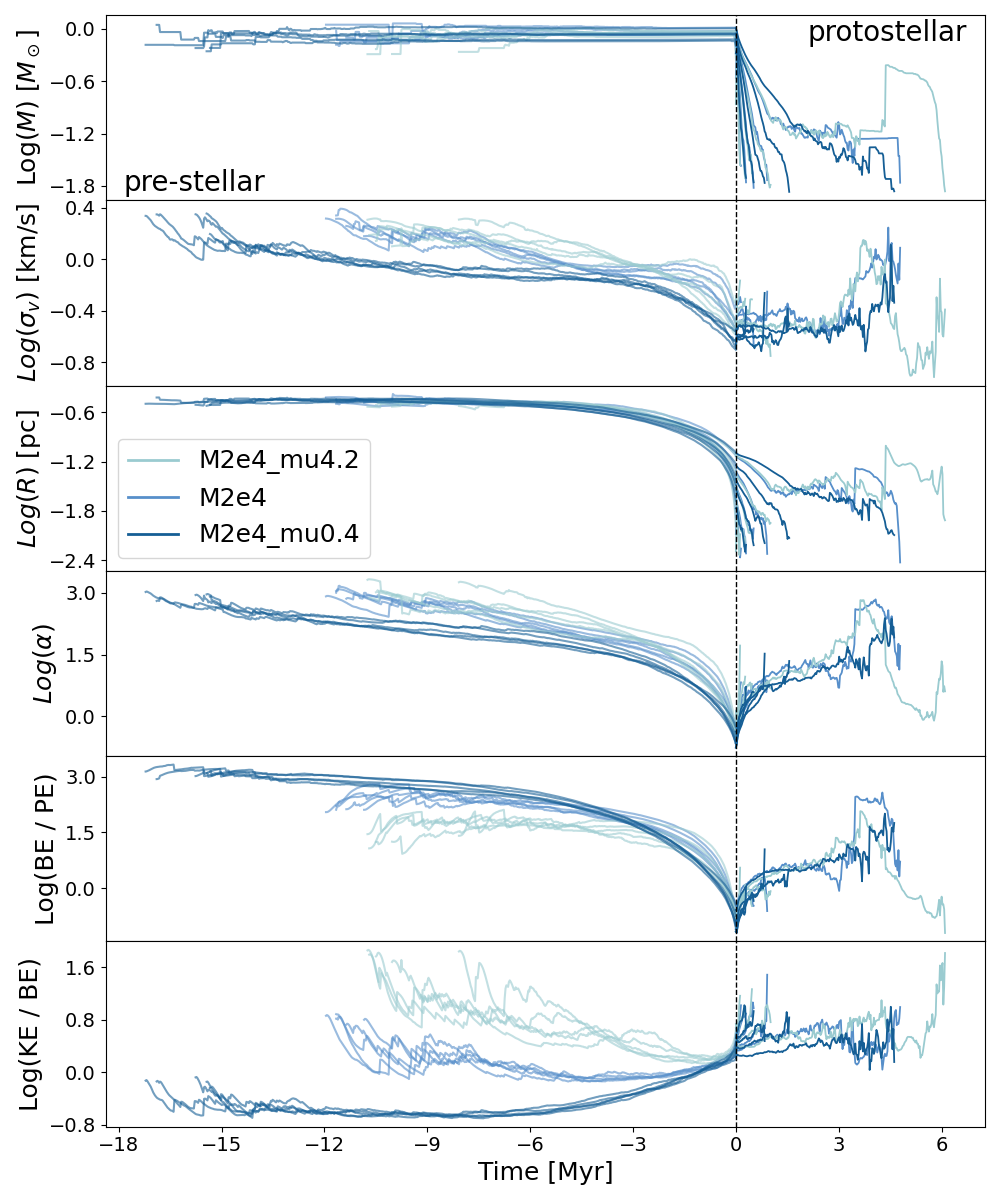}
    \caption{Time evolution of gas properties for the intermediate mass bin ($0.5<M<2M_\odot$), averaged across five bins sorted by protostellar duration. The three shades of lines correspond to the three simulation runs, with the darker lines indicating a stronger magnetic field. Panels show (top to bottom): gas mass, velocity dispersion, effective radius, virial parameter, BE/PE ratio, and KE/BE ratio. The vertical dotted line marks the onset of the protostellar phase. Low- and high-mass results are provided in the appendix.}
    \label{fig:prop_vs_time}
\end{figure}

Next we investigate the properties of the star-forming gas. Figure \ref{fig:prop_vs_time} shows the evolution of the gas properties of the intermediate mass stars, as defined in \S\ref{sec:properties}. The evolution for the low- and high-mass stars is shown in Appendix \ref{append:gas_properties}. 
For each cloud and mass bin, we divide the gas subsets by protostellar duration into five bins based on the formation time. 
The \med\ cloud has $551$ intermediate-mass stars, so each bin contains $\sim111$ stars, and the average protostellar durations for each bin are 0.08, 0.25, 0.42, 0.72, and 1.63 Myr. Note that short-time variation in the averages is produced by gas subsets being added or subtracted (due to completing accretion) from the bin.

\begin{figure}
    \centering
    \includegraphics[width=1.1\linewidth]{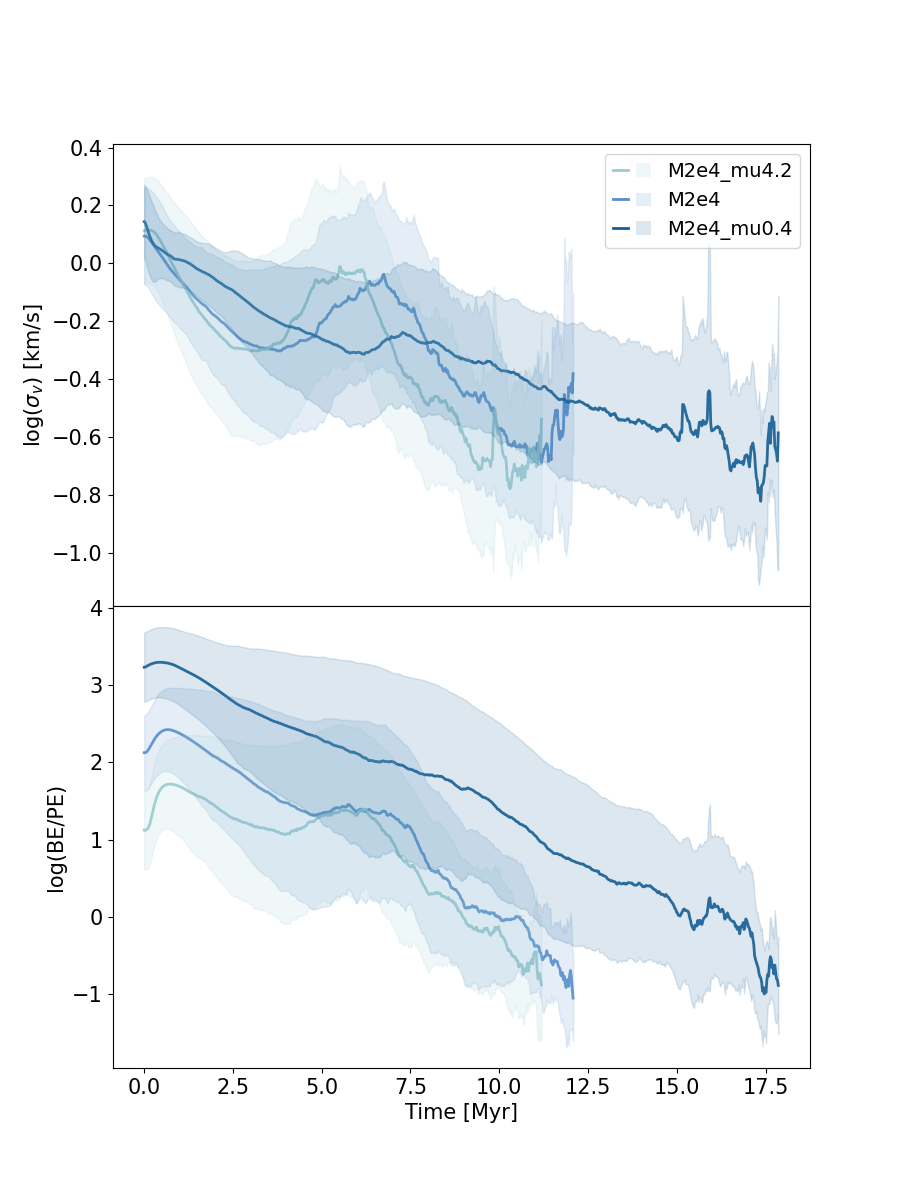}
    \caption{Average velocity dispersion and magnetic-to-gravitational energy ratio (BE/PE) across all cores versus time. 
    Solid lines indicate the mean value at each time, and the shaded regions show $\pm 1\sigma$ standard deviation. The three magnetic field strengths (\lo, \med, \hi) are indicated by the different colors.}
    \label{fig:ave_ratios}
\end{figure}

During the prestellar phase the mass for each individual subset remains constant. The lines in the top panel of Figure \ref{fig:prop_vs_time} show some small variation versus time since they represent averages over a number of subsets that have different prestellar (and protostellar) durations. Once accretion starts, the gas mass in most of the time bins declines rapidly. The longer-lived bins show more variation, since their accretion histories are also more complex (see \S\ref{sec:fitting}).  These persistent bins also tend to contain stars toward the higher end of the intermediate-mass distribution.

The second panel in Figure \ref{fig:prop_vs_time} shows that the prestellar velocity dispersions generally decline  with time; an initial slow decline caused by turbulence decay is partially offset 
by a combination of gravitational collapse of the cloud and increased turbulence driven by feedback from the earlier forming stars.  The velocity dispersion of the protostellar gas is relatively flat with some variation 
due to local motions driven by protostellar feedback. Again, the longer time bins, which contain more high-mass forming stars, exhibit the most variability.  

The impact of the cloud evolution and feedback on the star-forming gas is more clearly illustrated by the top panel of Figure \ref{fig:ave_ratios}, which shows the velocity dispersion versus time for all star-forming gas. Turbulence globally declines for the first $2-5$ Myr of the simulations, which is followed by a period of increased dispersion, during which the motions are driven by the global cloud collapse and vigorous stellar feedback. Due to strong magnetic support, run \hi\, has the longest pre-collapse period, the lowest star formation efficiency, and only a relatively mild enhancement in the velocity dispersion around 6 Myr.  At late times, the feedback suppresses star formation and the velocity dispersion continues to decline until the first supernovae occurs. 

The third panel in Figure \ref{fig:prop_vs_time} shows that the prestellar effective radius undergoes a slow contraction during the early prestellar phase, which accelerates as gravitational collapse occurs. The outer scale of the effective radius is set by the size of the simulated cloud, which is the same in all runs. The radii continue to contract during the protostellar phase, 
 with some variation due to the impact of feedback gas and averaging over histories with different durations. 

The fourth panel in Figure \ref{fig:prop_vs_time} shows that the virial parameter evolution follows that of the velocity dispersion: the initial star-forming gas $\alpha$ parameters are set by the cloud virial parameter and decline as the turbulence decays. Likewise, $\alpha$ increases during the protostellar phase as feedback drives turbulence locally and the size of the remaining gas reservoir shrinks.

The last two panels in Figure \ref{fig:prop_vs_time} show the evolution of the magnetic to potential energy ratio, BE/PE, and the kinetic to magnetic energy ratio, KE/BE. The magnetic ratio slowly declines as turbulence decays and the cloud globally contracts. Collapse roughly occurs when the magnitude of the gravitational potential energy exceeds the magnetic  and kinetic energy. This ratio increases during the protostellar phase as the magnetic field increases in the residual gas. The prestellar KE/BE ratio follows the velocity dispersion evolution: it is initially flat or declines with time. The ratio increases in the strong field run, \hi, as magnetic flux is removed from the gas.  
The KE/PE ratios of the protostellar gas increase slightly or remain flat as accretion begins.  
The bottom panel of Figure \ref{fig:ave_ratios}  shows that the KE/BE  ratio averaged overall all star-forming gas initially increases,   which is primarily due to the magnetic dynamo that increases the mean magnetic field strength \citep{GuszejnovGrudic2020a,LaneGrudic2022a}. As turbulence decays and the cloud collapses, the ratio declines. The two weaker field runs show another enhancement in the ratio of magnetic to potential energy during the epoch of peak star-formation activity, when vigorous stellar feedback causes the cloud to expand.

In all runs, the behavior of the protostellar gas is similar, indicating that local processes, namely stellar feedback, rather than global cloud parameters, drive the evolution. The evolution of the cloud properties is clearly visible in the trends in the prestellar gas; the higher magnetic field runs contract more slowly and begin collapse when the virial parameter is slightly lower, but otherwise the star-forming gas exhibits similar velocity dispersions, radii, and post-accretion evolution.

\begin{table}
\centering
\begin{tabular}{ccc}
\hline
Cloud Label & $p_{v}$ & $p_{m}$ \\
\hline
\lo & $0.52\pm0.02$ & $1.14\pm0.02$ \\
\med & $0.55\pm0.01$ & $1.12\pm0.02$ \\
\hi & $0.47\pm0.01$ & $0.98\pm0.01$ \\
\hline
\end{tabular}
\caption{Best-fit slopes and uncertainties for the linear least squares fits shown in Figure \ref{fig:proto-onset}. $p_v$ is the power-law slope of velocity dispersion versus radius ($\sigma_v\propto{r^{p_v}}$), and $p_m$ is the power-law slope of mass versus radius ($M\propto{r^{p_m}}$).}
\label{fits}
\end{table}

\begin{table}
\centering
\begin{tabular}{cccc}
\hline
Cloud Label & $M [M_{\odot}]$ & $\sigma_v[\text{km/s}]$ & $r[\text{pc}]$\\
\hline
\lo & $0.45^{+0.59}_{-0.20}$ & $0.20^{+0.13}_{-0.07}$ & $0.02^{+0.02}_{-0.01}$ \\
\med & $0.44^{+0.63}_{-0.19}$ & $0.21^{+0.15}_{-0.08}$ & $0.02^{+0.02}_{-0.01}$ \\
\hi & $0.34^{+0.33}_{-0.12}$ & $0.16^{+0.08}_{-0.05}$ & $0.02^{+0.02}_{-0.01}$ \\
\hline
\end{tabular}
\caption{Median values of the protostellar onset properties for the mass, velocity dispersion, and radius with 25\% and 75\% quartiles indicated.}
\label{tab:medians}
\end{table}

\begin{figure}
    \centering
    \includegraphics[width=1\linewidth]{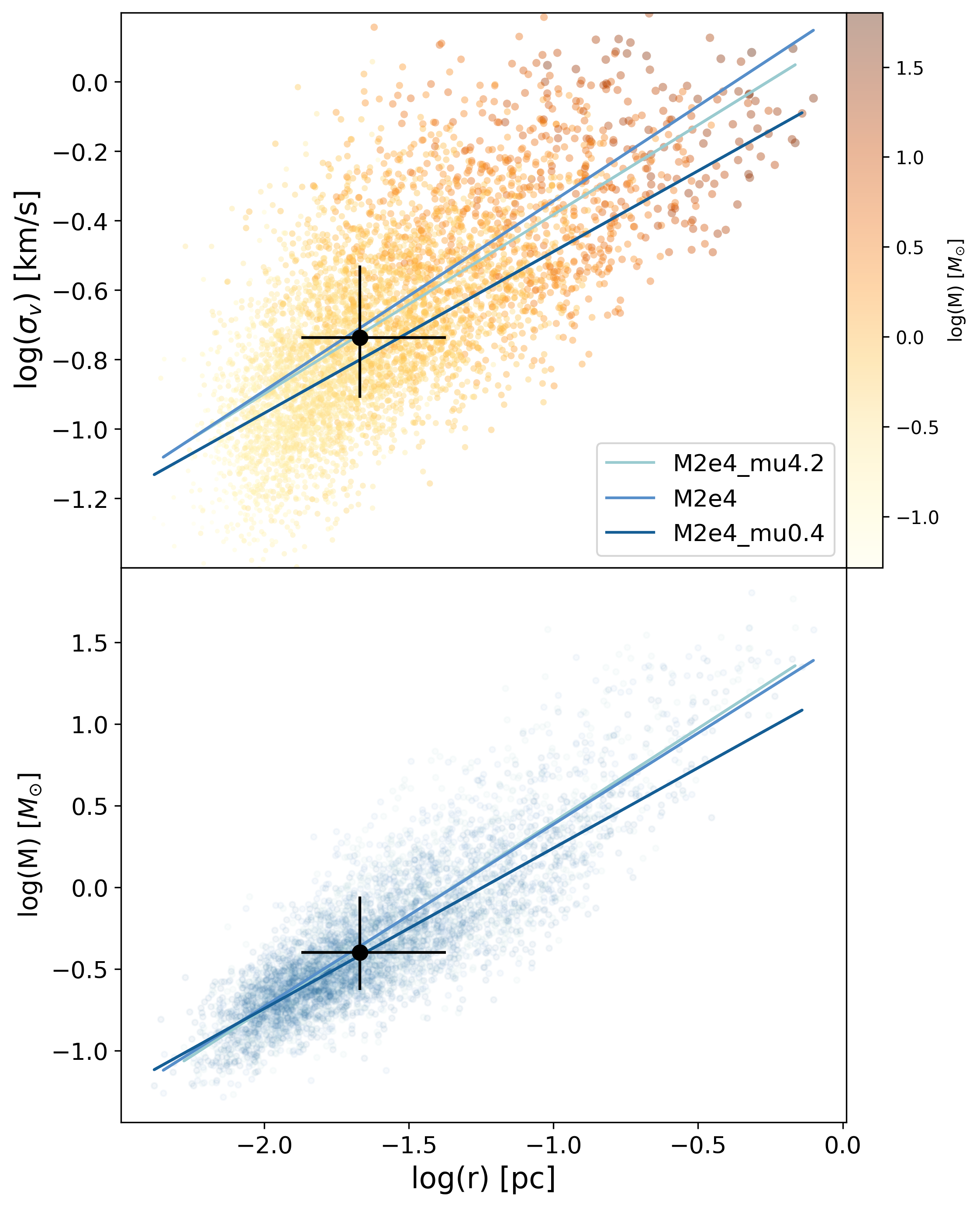}
    \caption{Log velocity dispersion  (top) and log mass (bottom) versus log effective radius  at the onset of accretion (start of the protostellar phase) for all stars in each simulation. In the top panel the points are colored by mass. Solid lines represent the best-fit linear regressions for each simulation, colored by magnetic field strength. The median value is marked in black. The fit parameters are reported in Table \ref{fits}. }
    \label{fig:proto-onset}
\end{figure}

At the time the protostar forms in each gas subset, the gas distributions, while not cores per se, exhibit the similar correlations between mass, radius, and linewidth that are characteristic of observed cores and clouds. Figure \ref{fig:proto-onset} shows that this star-forming gas follows well-defined linewidth-size relations ($\sigma_{v}\propto{r}^{p_v}$) and mass-size relations ($M\propto{r}^{p_m}$). The velocity-radius slope, $p_v$, spans 0.47 to 0.55. These values are consistent with supersonic turbulence driving gas motions, which produce $p_v \simeq 0.5$  \citep{Larson1981c}. The mass-radius slope, $p_m$, ranges from 0.98 to 1.14, which suggests that the gas distributions resemble Bonnor-Ebert spheres, where the density profile $\rho \propto r^{-k_\rho}$ with $k_\rho = 2$ and gravity and thermal pressure are comparable.

The gas in all three simulations exhibits similar correlations, with the strong magnetic field run exhibiting slightly flatter slopes. This is likely because these gas subsets are less turbulent as shown in Figure \ref{fig:prop_vs_time}.

\subsection{Accretion History}\label{sec:fitting}

The rate of change of the amount of remaining gas indicates the accretion rate of a given protostar. Several star formation theories, such as the isothermal sphere \citep{Shu1977a}, turbulent core \citep{McKeeTan2002a}, and competitive accretion \citep{Bonnell1994a} models, predict how accretion varies as a function of time and stellar mass. Thus, how quickly the star-forming gas accretes  provides clues about the underlying star-formation paradigm in the simulations.

We fit the accretion history of each star with a power-law model based on the functional form derived in \citet{McKee2010}. In this framework, the remaining mass, $m_r$, available for accretion evolves as
\begin{equation}
{m_r(t)}={\left(1-{\frac{t}{t_f}}\right)^{\frac{1}{1-j}}m_f},\label{eq:accretion}
\end{equation}
where $m_f$ is the final stellar mass, i.e., the stellar mass after accretion ends, $t_f$ is the accretion duration for that star, and $j$ is a free parameter that varies between 0 and 1 that controls the time-dependence of the accretion rate. This formulation captures a range of accretion behaviors.  A value of $j=0$ corresponds to a collapsing isothermal sphere, where the accretion rate is constant in time and independent of stellar mass  and where the host core has a density profile with  $k_\rho = 2$ \citep{Shu1977a}. A value of $j={1/2}$ corresponds to the turbulent core model  for a core with $k_\rho= 3/2$ \citep{McKeeTan2003a}. In this case, the accretion rate increases with the final stellar mass,  $\dot m \propto m_f^{1/4}$, such that more massive stars accrete at higher rates.   In general, $j$ can be written in terms of the gas density profile as \citep{McKeeOffner2010a}
\begin{equation}
 j = \frac{3 (2 - k_\rho)}{2 (3-k_\rho)}.    
\end{equation}
Consequently, higher (lower) values of $j$ can be interpreted as accretion from a gas reservoir with a shallower (steeper) density profile.
In the competitive accretion scenario, the accretion rate is driven by tidal effects and $\dot m 
\propto m ^{2/3}$ \citep{BonnellBate2001a}. This corresponds to a value of  $j=2/3$ in the \citet{McKeeOffner2010a} formalism.
When $j > 0$, as in the turbulent core and competitive accretion cases, the accretion rate increases as the protostellar mass, $m$, approaches its final value of $m_f$.

\begin{figure}
    \centering
    \includegraphics[width=1\linewidth]{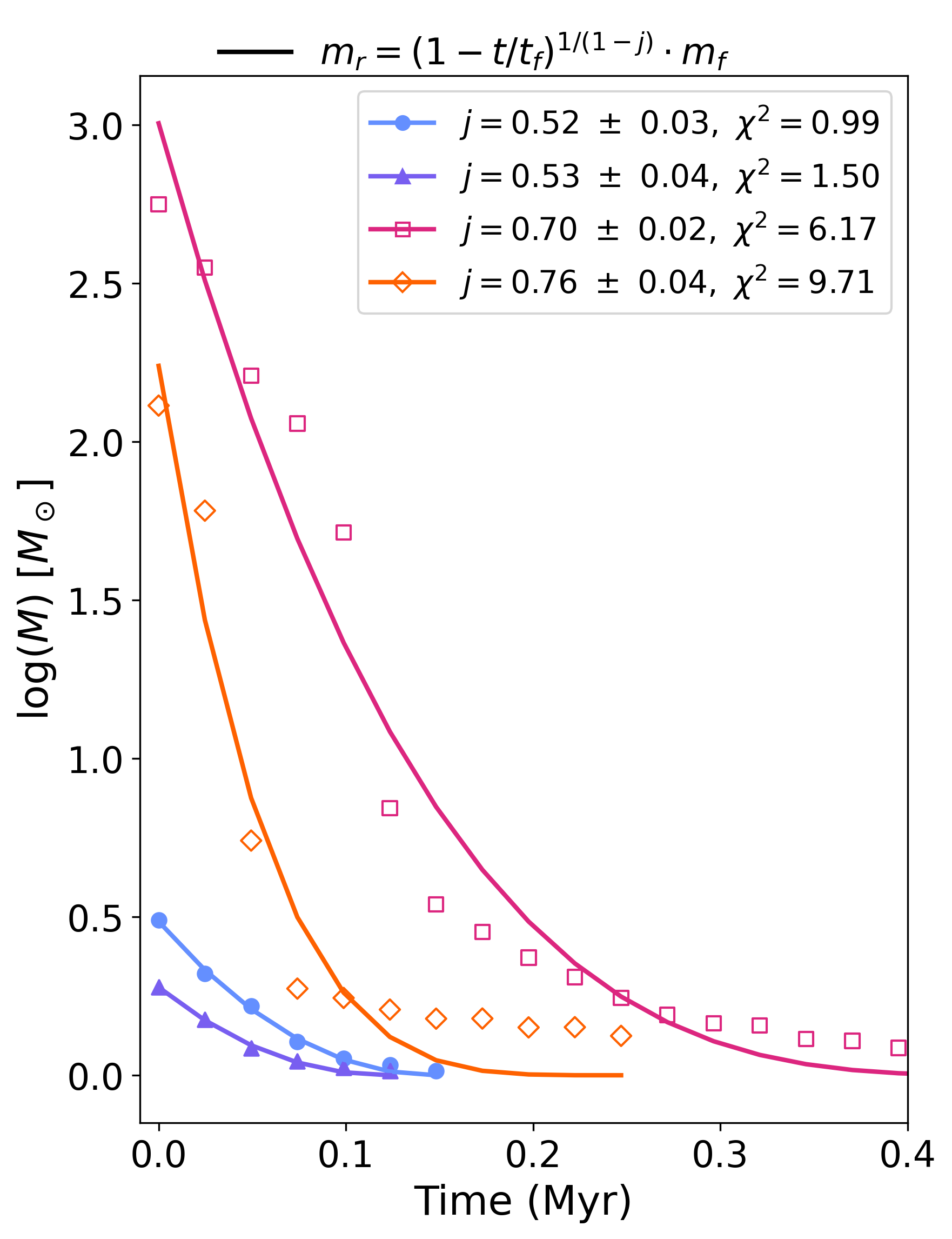}
    \caption{Gas mass versus time for four typical stars, where the filled symbols represent cases well-fit by Equation \ref{eq:accretion} and the open symbols represent cases with poor fits. The latter cases have significantly variable accretion.}
    \label{fig:example-fits}
\end{figure}

Figure \ref{fig:example-fits} shows accretion fits for four typical stars. The stars exhibit a range of accretion histories, from steep, early growth to slow, extended accretion. The best-fit values of $j$ and the associated $\chi^{2}$ goodness-of-fit metrics are indicated in the legend. We find that there are two classes of accretion behaviors. Well-behaved accretors experience smooth accretion that is well-fit by a single power-law as in Equation \ref{eq:accretion}. Variable accretors undergo abrupt changes in their accretion; these protostars are not well-fit by Equation \ref{eq:accretion}. The chi-squared value is a good indicator of the degree of variability in a given star's accretion history.

We find that the quality of fit is generally better for lower-mass stars. These stars tend to complete accretion more rapidly and exhibit smoother, monotonic growth histories that align well with the continuous power-law functional form of the model (see the curves with filled symbols in Figure \ref{fig:example-fits}). In contrast, higher-mass stars typically accrete for longer durations, during which their accretion often exhibits a ``stair-step" behavior: periods of rapid accretion interspersed with plateaus during which the star accretes relatively little (open symbols in Figure \ref{fig:example-fits}).

\begin{figure*}
    \centering
    \includegraphics[width=1\linewidth]{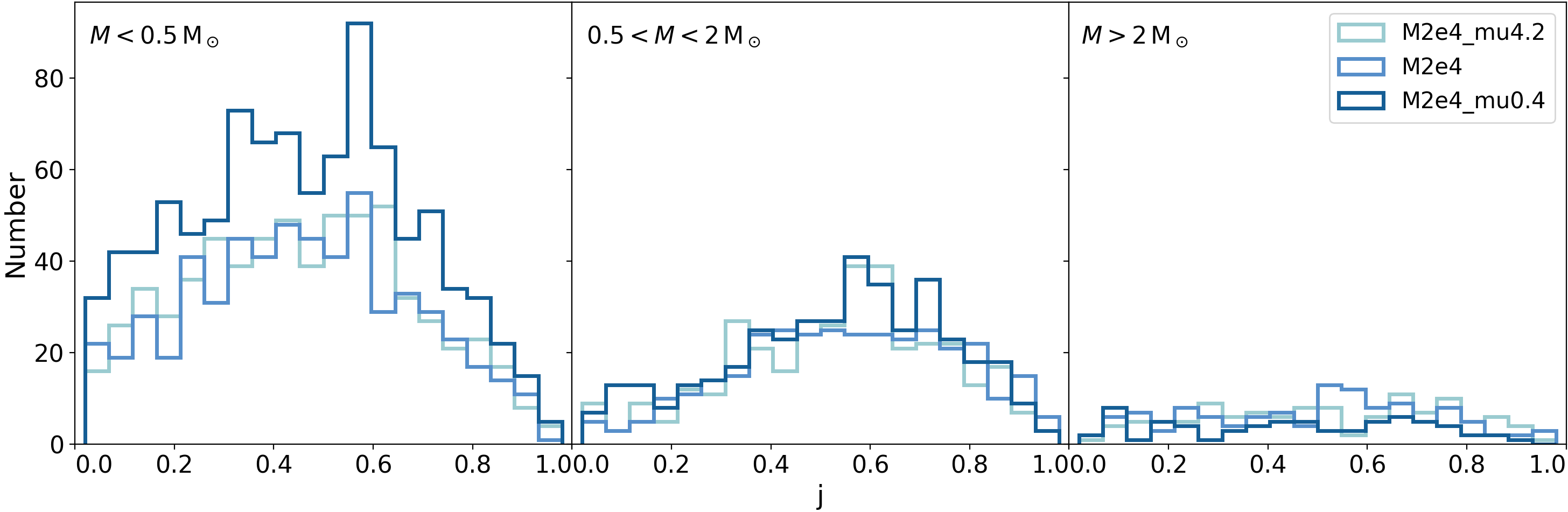}
    \caption{Distribution of $j=1-1/b$ parameter values for low- ($M<0.5M_{\odot}$), intermediate- ($0.5<M<2M_{\odot}$), and high-mass ($M>2M_{\odot}$) stars across the three clouds. A $\chi^2$ cutoff of 3.0 was applied to exclude poorly fit mass accretion histories, and the resulting distributions peak near $j\approx0.5$.}
    \label{j-values}
\end{figure*}

Figure \ref{j-values} displays the distribution of best-fit $j$ values for all  well-behaved accretors, binned by final stellar mass and initial cloud magnetic field strength. We define stars as well-behaved if $\chi^{2} < 3$  and exclude stars with poor fits ($\chi^{2}\ge 3$). This threshold visually corresponds to the division between apparently good and bad fits (e.g., Figure \ref{fig:example-fits}).  Most stars are well-fit by Equation \ref{eq:accretion}. Across all simulations, this criterion excludes $\sim 8\%$ of the low-mass stars, $\sim16\%$ of the medium-mass stars, and $\sim 22\%$ of the high-mass stars. 
We find that the resulting distributions have relatively similar shapes and peaks for choices of $\chi^{2}=2$ or 4, so our results are not sensitive to this specific choice.

Figure \ref{j-values} shows that all runs form stars with a broad spread in $j$ values that range from 0 to $\sim 0.9$; however, the $j$ distribution changes with mass.   Table \ref{tab:jvals} gives the median $j$ values for each mass bin for each simulation. The lowest mass stars have a distribution peaked around $j \sim 0.4$, while the intermediate-mass stars have a distribution skewed towards slightly higher $j$ values with a peak around $0.5$,   consistent with the turbulent core model. Meanwhile, the higher mass stars have a flat distribution of $j$ values. They do not strongly favor  any $j$ value. This may be in part because there are relatively fewer of them and because, even for lower $\chi^2$ fits, there is some degree of variation in their accretion history.

  As discussed, variation in $j$ can be driven by variation in the gas profile. However, there are other factors as well.  Intermediate values of $j$, e.g., $j\sim 0.2$ or $j \sim 0.75$, could be caused by accretion that is not fully described by any of the three models but rather accretion that is a composite of different accretion types.  \citet{McKeeOffner2010a} also defined ``two component" accretion models, where accretion is partially isothermal and partially turbulent or competitive. Accretion can also decline at late times and ``taper" exponentially \citep{McKeeOffner2010a,HartmannHerczeg2016a}. A numerical study by \citet{OffnerChaban2017a} found that protostellar accretion in isolated turbulent, magnetized cores is well-fit by a tapered turbulent accretion model. These models can all be described by smooth accretion histories but ones with additional model parameters.
Variation in the model coefficients with time, such as due to changing gas temperature and surface density, would also impact the fitted $j$. Indeed, a range of accretion modes is the most likely outcome in a messy star-forming environment where protostars do not accrete from clearly-defined, isolated, uniform temperature cores.
 
Interestingly, for all masses the distributions of fitted $j$ appear insensitive to the cloud magnetic field strength. This suggests that protostellar mass assembly proceeds through gravitationally regulated accretion, with turbulence and dynamics, rather than magnetic fields, playing dominant roles in shaping the accretion flow.

\begin{table}
\centering
\begin{tabular}{lccc}
\hline
Cloud Label & $M<0.5M_{\odot}$ & $0.5<M<2M_{\odot}$ & $M>2M_{\odot}$ \\
\hline
\lo  & $0.407^{+0.151}_{-0.148}$ & $0.493^{+0.127}_{-0.160}$ & $0.397^{+0.159}_{-0.151}$ \\
\med  & $0.420^{+0.156}_{-0.151}$ & $0.513^{+0.111}_{-0.171}$ & $0.412^{+0.197}_{-0.143}$ \\
\hi & $0.422^{+0.156}_{-0.155}$ & $0.509^{+0.131}_{-0.170}$ & $0.441^{+0.173}_{-0.228}$ \\
\hline
\end{tabular}
\caption{Median $j$ values with interquartile ranges for each cloud and mass category represented in Figure \ref{j-values} at $\chi^2$ cutoff = 3.}
\label{tab:jvals}
\end{table}

\section{Discussion} \label{discussion}

\subsection{The Properties of Cores versus the Properties of Star-Forming Gas}

In contrast to the STARFORGE study by \citet{OffnerTaylor2025a}, which analyzed the same three simulations, the present work does not define the cores embedding each protostar but instead focuses only on the subset of gas that actually accretes. These two structures -- cores and accreting gas sets -- naturally share material and overlap most directly during the protostellar phase. As Figures \ref{fig:ex_reservoir_pre} and \ref{fig:ex_reservoir_post} show, at certain times the gas subsets appear as cohesive structures with sizes of $\sim 0.1$\,pc and resemble cores in terms of their size and shape. This is especially true in the case of low-mass stars, which have relatively short accretion times and the accreting gas is relatively local to the protostar at the time of its formation. However, considering that a significant portion of an identified core will eventually be dispersed by stellar feedback \citep[e.g.,][]{MatznerMcKee2000a, MachidaHosokawa2013a,OffnerArce2014a,OffnerChaban2017a} and that cores, by definition, are circumscribed by some density-based (or flux-based) boundary, the degree of similarity between the properties of cores and gas subsets is somewhat surprising. 

Comparing the properties of the gas subsets at the time of protostar formation to those of the cores identified in \citet{OffnerTaylor2025a} indicates that both populations follow a mass-size relation $M \propto R^1$, independent of the cloud field strength. Similarly, both obey linewidth-size relations, although the relation for the cores identified in \citet{OffnerTaylor2025a} is flatter, with $\sigma \propto R^{0.2-0.3}$. This may be because the cores sample a range of times with respect to protostar formation, whereas the gas subset properties are captured at the point of protostar formation when the gas kinematics are less influenced by stellar feedback \citep[e.g.,][]{NeralwarColombo2024a} and likely still reflect the larger turbulent cascade. The gas-subset relations are consistent with the observed linewidth-size \citep{Larson1981c, CaselliMyers1995a,ColomboRosolowsky2019a,ChoudhuryPineda2021a,NeralwarColombo2022b} and mass-size relations \citep{NeralwarColombo2022a} for cores and clumps,  although there is a large variation between low- and high-mass regions and with different observational tracers.

At the time of protostar formation, the properties of the gas subsets, simulated cores, and observed cores are also similar. The median radius, velocity dispersion, and mass of the starless cores in \citet{OffnerTaylor2025a} are $\sim 0.08$\,pc, $\sim 0.3$ km/s, and $\sim 0.4-0.5 M_\odot$, respectively. Table \ref{tab:medians} shows that the gas subsets have similar median masses and velocity dispersions of $\sim 0.3-0.4 M_\odot$ and $\sim 0.2$ km/s. The median gas subset radii of $\sim 0.03$ pc are more compact than those of the starless cores but is comparable to the radii of the STARFORGE protostellar cores, which are $\sim 0.03-0.04$ pc. The simulated star-forming gas properties are also comparable to those of dense cores observed in NH$_3$, although we stress that these values are sensitive to the molecular tracer and the core definition \citep{KirkFriesen2017a,KerrKirk2019a,ChenPineda2019a,OffnerTaylor2022a}. 
Altogether, these similarities suggest that the properties of observed cores reflect those of the actual star-forming gas, despite the many differences between observational surveys and simulation approaches. The accreting gas properties, which are relatively insensitive to the cloud magnetic field strength, also underscore that turbulence (early) and stellar feedback (later) are largely responsible for regulating the evolution of both cores and star-forming gas.

\subsection{Comparison of High-Mass and Low-mass Star Formation}

Our analysis highlights some key similarities and differences between the formation of low-mass and high-mass protostars.
While the gas subsets of forming high-mass stars are initially more extended (c.f.~Fig.~\ref{fig:ex_reservoir_pre} and \ref{fig:ex_reservoir_post}) and accretion occurs over a longer time period (see Table \ref{tab:lifetimes}), the gas in both cases follows similar relations as shown in Figure \ref{fig:proto-onset}. This suggests that the lack of high-mass cores in observations \citep[e.g.,][]{LouvetNaupane2019a,SvobodaShirley2019a,BarnesHenshaw2021a,MoriiSanhueza2023a} may not necessarily be because the gas is not present locally but rather because it is challenging to define a core boundary a priori that encompasses most of the accreting gas. \citet{PelkonenPadoan2021a} suggest that high-mass cores do not exist in part because gas continues to in-flow from larger distances and accrete along filaments. Here we find that the accreting gas in STARFORGE is not necessarily distributed in a filamentary configuration at the time of protostar formation (c.f., Figure \ref{fig:ex_reservoir_pre} and \ref{fig:ex_reservoir_post}), but it does accrete from larger distances than the gas accreting onto low-mass protostars.   It is not that the gas properties a priori determine final stellar masses,  but rather that certain gas distributions produce higher mass stars because feedback or turbulent intermittency does not halt accretion or disrupt collapse.  Appendix \ref{append:gas_dist} shows additional examples of the accreting gas distributions for forming low- and high-mass stars. 

While the low- and high-mass star-forming gas properties satisfy the same property relations as shown in Figure \ref{fig:proto-onset}, the time-evolution indicates that the gas accretes differently. Low-mass protostars are relatively continuous accretors, whereas high-mass protostars experience more variable accretion. Note that this does not mean that episodic accretion, brief periods during which the accretion rate increases by a factor of $\sim$10-100 \citep{HartmannKenyon1996a,AudardAbraham2014a}, does not occur.  Neither the spatial resolution nor snapshot spacing is sufficiently high to capture such bursts and, regardless, the incremental change in mass would be small compared to the total. Rather, the accretion histories show periods in which the accretion rate  precipitously declines or increases on $\sim 0.05$~Myr timescales (see Fig.~\ref{fig:example-fits}).

We find that the accretion histories are generally well fit by the \citet{McKeeOffner2010a} formalism with a broad range of $j$ values, suggesting that the accretion histories are described by a mix of isothermal sphere, turbulent core, and competitive accretion with gas profiles of $k_\rho \sim 1.5-2$. This is consistent with our fitted  $M\propto R^1$ scaling relation. These profiles are in agreement with those of observed cores, where the profiles range from  $k_\rho \sim 2$ for low-mass cores to $k_\rho \sim 3/2$  for higher-mass cores \citep{McKeeOstriker2007a}.
This behavior is consistent with the three-stage collapse framework described by \citet{CollinsLe2024a}, where higher mass cores exhibit bursts of accretion until the bulk of the mass is accreted during a ``singularity" phase. In our simulations, the accretion is also mediated by stellar feedback.
While the spread in the fit exponents may be due to hybrid accretion, tapered accretion or changing physical conditions, the histories are sufficiently variable in the high-mass case that no smooth accretion model provides a good fit in most cases. This accreting gas is both more turbulent and subject to stronger stellar feedback \citep{NeralwarColombo2024a}. In addition, high-mass protostars are also more likely to be multiples \citep{OffnerMoe2023a,GuszejnovRaju2023a} and thus more strongly impacted by dynamical interactions.  \citet{GenerozovOffner2025a} find that higher mass stars undergo many more encounters and exchange companions more frequently than low-mass stars do; these events are likely to disrupt accretion. This reflects the more complex birth environment of high-mass stars compared to that of lower mass stars.  

\subsection{Comparison to Prior Numerical Simulations}

A number of prior numerical studies have investigated star formation by following the gas evolution using Lagrangian particles. These approaches either used smoothed particle hydrodynamics, which tracks individual gas parcels similarly to the meshless finite mass method employed by {\it Gizmo}, or used massless Lagrangian tracer particles that effectively sample the flow.  In the former approach all star-forming gas can be followed exactly, while in the latter case the tracers provide a statistical representation of the underlying dynamics.

Several of these previous works also found that core properties are largely independent of the parent cloud properties. \citet{MoczBurkhart2018a} used tracer particles to examine the formation of dense cores in post-shock material. While the strength of the magnetic field dictated the degree of alignment between the field and shock orientation, the core profiles were independent of the cloud magnetic field strength.   Similarly, \citet{KuznetsovaHartmann2019a} used tracer particles to follow the origin of protostellar angular momentum, which they found was independent of the rotation of the parent cloud. These studies suggest that compact dense regions are decoupled from the larger cloud properties and evolve independently. This occurs if cores are regulated by their own self-gravity and by turbulence, which is self-similar. 

Several other studies have investigated the evolution of the star-forming gas prior to its accretion, finding that the gas distributions typically extend beyond the identifiable core region \citep{BonnellVine2004a,SmithLongmore2009a,Pelkonen2021,CollinsLe2023a}.
\citet{CollinsLe2023a} conducted a detailed analysis of  pre-collapse gas, finding that the distributions are heterogeneous and that gas forming different stars often occupies overlapping volumes. Similarly \citet{Pelkonen2021} showed that high-mass stars accrete from much larger volumes than those of their host cores, which they define as gravitationally bound regions around the forming protostars. In other words, stars accrete a significant amount of material that is not bound at the time of protostar formation.  \citet{Pelkonen2021} find that the discrepancy between the core gas and the accreting material increases with final stellar mass, such that only $\sim 10$\% of the accreting material is included in the initial bound core for higher mass stars. 
With the exception of STARFORGE, these calculations did not include stellar feedback, which may halt the accretion of unbound material. However, \citet{GenerozovOffner2025a} find that the gas in STARFORGE exhibits similar behavior, where the cores containing forming high-mass stars are less likely to be gravitationally bound at the time the protostar forms. 

These Lagrangian analyses are consistent with studies showing that stellar masses are not well correlated with the masses of their host cores, even when the two mass functions have similar shapes \citep[e.g.,][]{SmithClark2009a,OffnerClark2014a,SmullenKratter2020a}.  If much of the accreting gas is not contained in the parent dense core then, by definition, it is unsurprising that core masses and final protostellar masses are poorly correlated.

\section{Conclusion} \label{conclusion}

We present a Lagrangian 
analysis of the evolution of 
star-forming gas in the STARFORGE simulations, focusing on three GMCs with identical initial conditions but varying magnetic field strengths. By isolating and tracking the gas that accretes onto individual stars, we characterize how key properties like mass, velocity dispersion, radius, and energy evolve before and after each protostar forms.
We also evaluate the impact of global cloud conditions versus local processes such as 
stellar feedback on star formation in these GMCs.

We find that the properties of the prestellar gas before protostar formation are influenced by the global cloud properties. For example, 
 star formation starts later in the stronger magnetic field simulations, reflecting the role of magnetic support in 
delaying cloud collapse. Likewise, the gas velocity dispersion and energies follow those of the parent cloud.
In contrast, once a protostar forms, the lifetime of the accreting gas shows only a weak dependence on the magnetic field strength and instead scales strongly with stellar mass. On average, low-mass stars complete accretion within $\sim$ 0.2 Myr, while high-mass stars continue accreting for up to 0.8 (\lo) Myr  or 1.7 (\hi) Myr.

At the time of protostar formation, the star-forming gas exhibits linewidth-size and mass-size correlations consistent with classical turbulence-regulated scaling relations ($\sigma \propto R^{0.5}$ and $M \propto R^1$), despite some gas distributions not resembling compact or centrally concentrated cores. These relations are similar across the three magnetic field strength runs, although the slopes are slightly shallower in the more strongly magnetized cloud.

By fitting accretion histories with a time-dependent power-law model, we uncover two broad classes of accretion behavior: low-mass stars typically show smooth accretion histories consistent with isothermal sphere or turbulent core models, whereas higher-mass stars exhibit more variable, stair-step growth.  While all three models -- isothermal sphere, turbulent core, and competitive accretion -- provide good fits to some accretion histories, no one model provides a full description for all star masses.  The accretion trends are also largely insensitive to the global magnetic environment, suggesting that local turbulence, stellar feedback, and dynamics play the dominant role in shaping gas accretion and dispersal.

These results highlight the importance of directly tracing star-forming gas to capture the full complexity of its evolution. 
Our findings also emphasize that star-forming gas can exhibit observationally core-like properties without necessarily resembling individual, cohesive cores. 

\begin{acknowledgements}
The authors acknowledge helpful comments from Aleksey Generozov, Kaitlin Kratter,  Claude-Andr\'e Faucher-Gigu\`ere, and an anonymous referee.
SSRO and AK were supported by NSF AST-2107340. SO also acknowledges support from NSF AST-2107942, NASA 80NSSC23K047, NSF AST-2407522, a Peter O'Donnell Distinguished Researcher Fellowship, and a Donald Harrington Fellowship. NF was supported by a DOE Computational Science Graduate Fellowship. The Flatiron Institute is a division of the Simons Foundation.
The analysis was enabled by the Frontera computing project at the Texas Advanced Computing Center. Frontera is made possible by the NSF award OAC-1818253.
\end{acknowledgements}

\bibliographystyle{aasjournalv7}
\bibliography{references.bib,biblio_complete_2023.bib}

\appendix

\section{Properties Time Evolution}\label{append:gas_properties}

Time evolution of the gas properties for low-mass and high-mass stellar populations. The property progressions are similar to those of the intermediate-mass stars. Likewise, within each class there is diversity in the protostellar phase in terms of both the time-dependent behavior and the length of accretion times.  

\begin{figure}[h]
    \centering
    \includegraphics[width=0.49\linewidth]{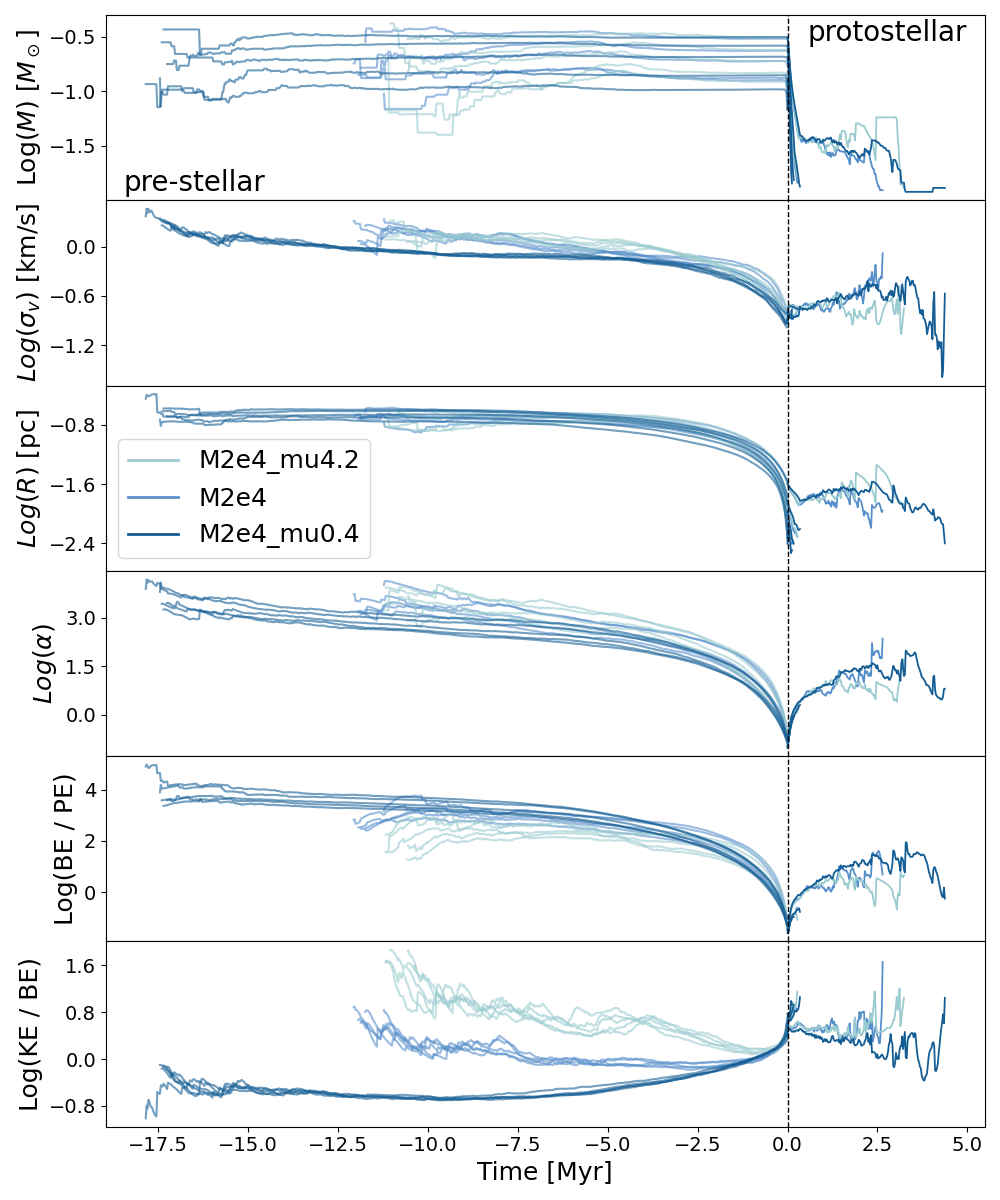}
    \includegraphics[width=0.49\linewidth]{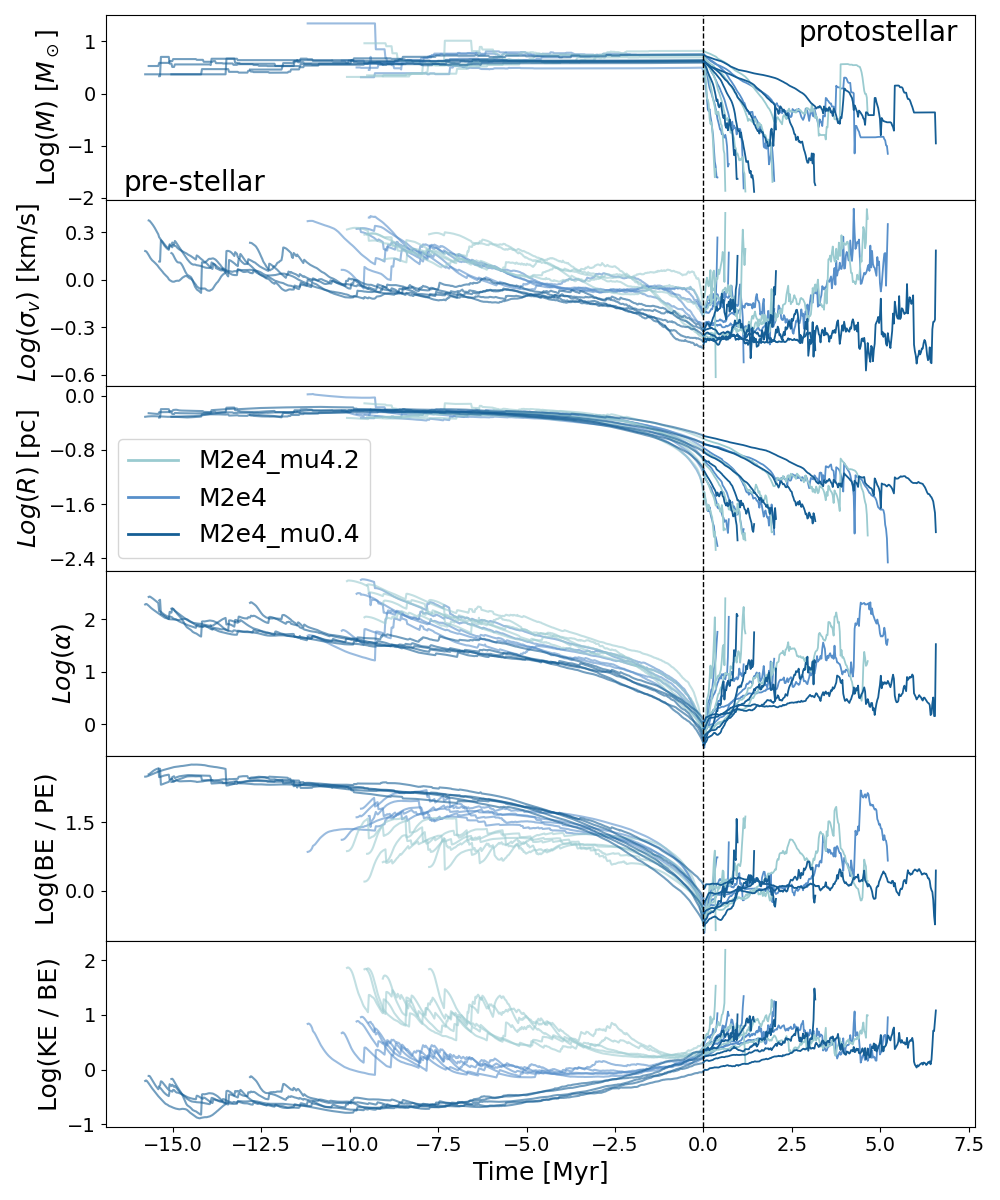}
    \caption{Same as Figure \ref{fig:prop_vs_time}  but for gas accreting onto the low-mass stars (left) and high-mass stars (right).
  There are $1494$ low-mass stars, so each bin contains $\sim299$ stars and the average protostellar durations of the bins are 0.03, 0.03, 0.12, 0.22, and 0.61 Myr. There are  $204$ high-mass stars; the bins each contain $\sim41$ stars and the average protostellar durations are 0.26, 0.59, 0.91, 1.54, and 2.71 Myr.    
    }
    \label{fig:enter-label}
\end{figure}

\section{Gas Distributions}\label{append:gas_dist}

\begin{figure}
    \centering
    \includegraphics[width=0.9\linewidth]{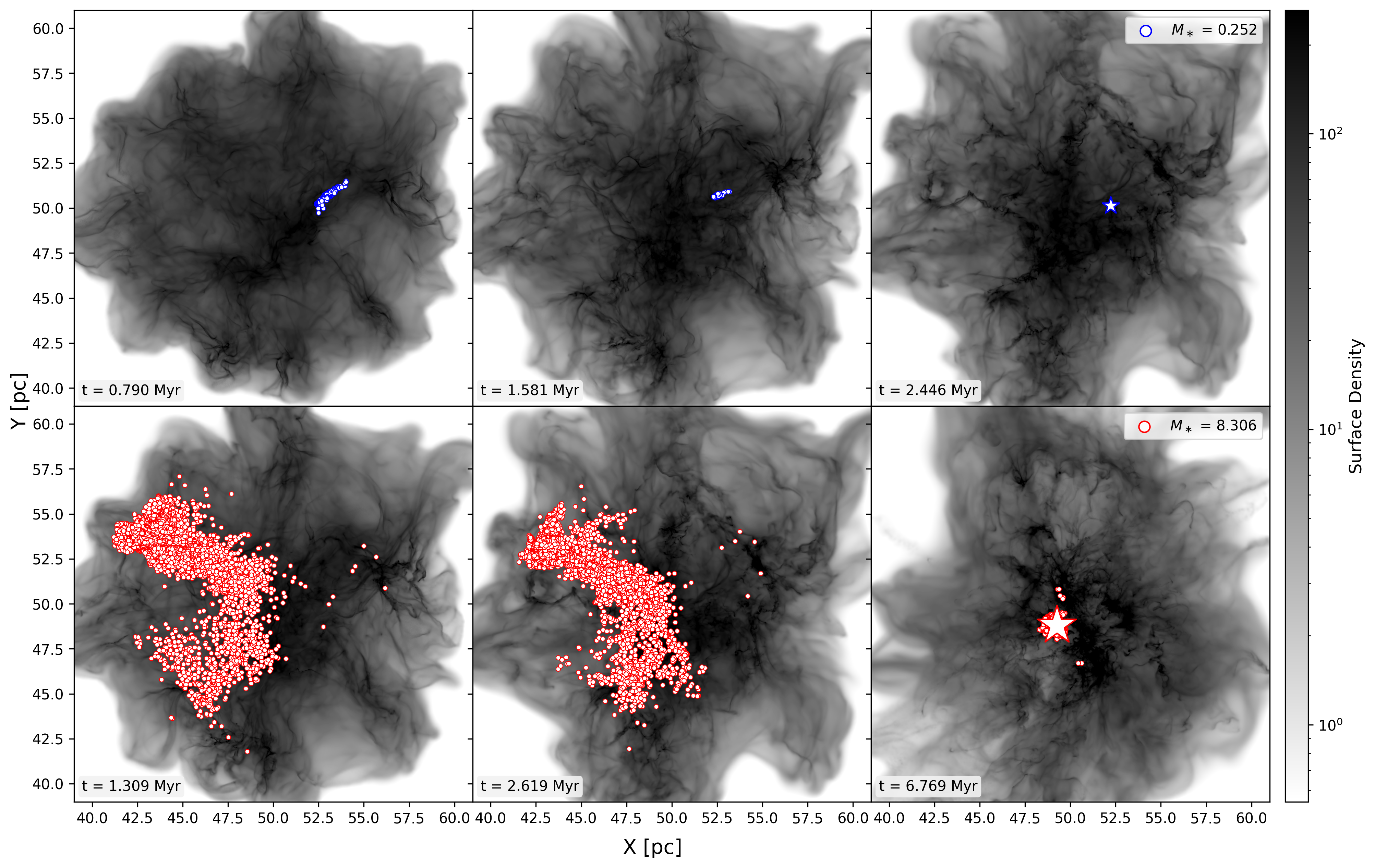}
    \includegraphics[width=0.9\linewidth]{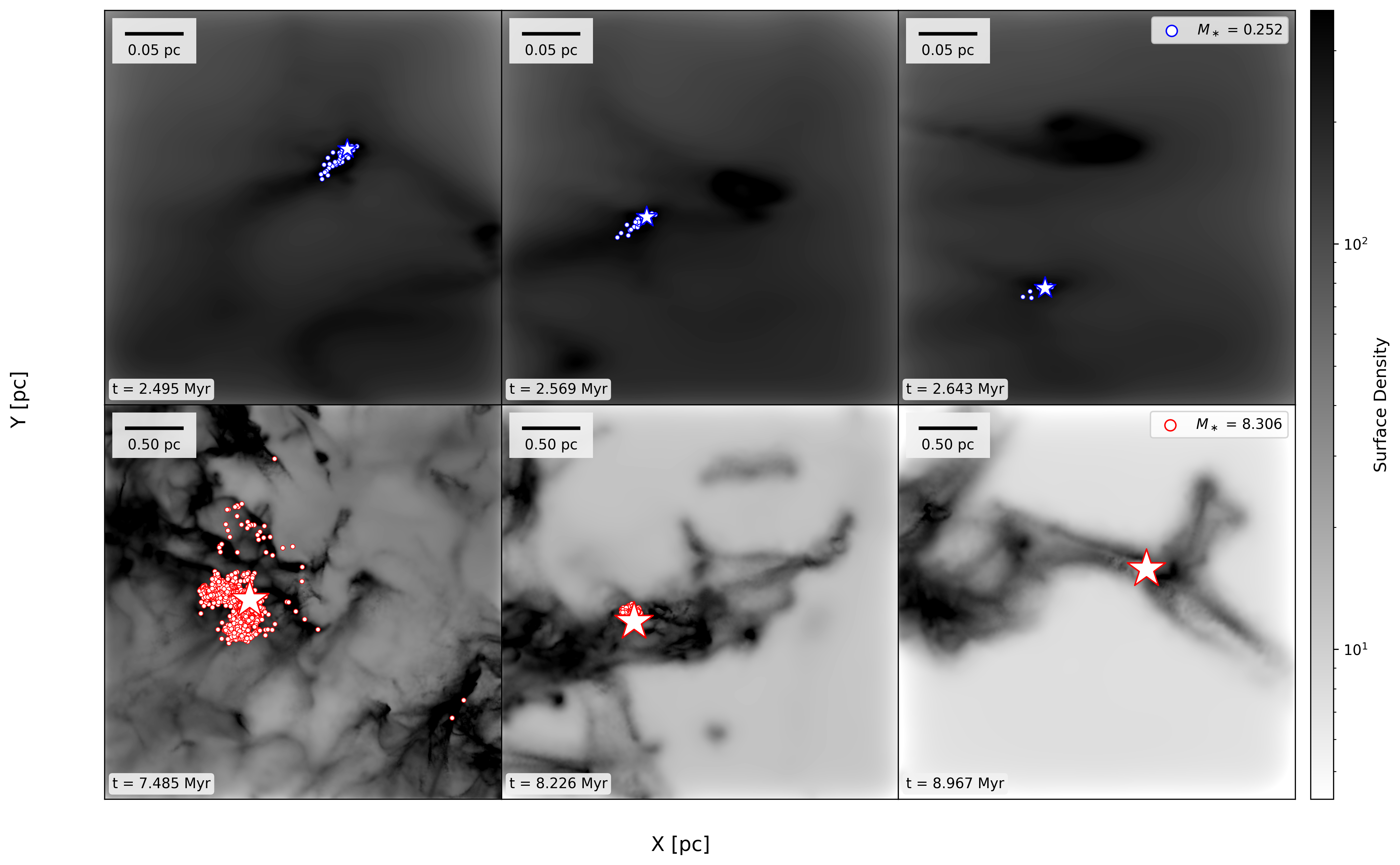}
    \caption{Top panels: Same as Figure \ref{fig:ex_reservoir_pre} but for stars with final masses $0.252\text{ }M_{\odot}$ (top) and for $8.306\text{ }M_{\odot}$ (bottom). Bottom panels: Same as  Figure \ref{fig:ex_reservoir_post} but for the two stars shown in the top panels. The stars are forming in run \med. \label{fig:gas1}}
\end{figure}

Examples of the prestellar and protostellar gas distributions for several additional protostars.  The high-mass prestellar distribution morphologies are somewhat heterogeneous with Figure \ref{fig:gas1} appearing more spherical and compact, while the high-mass prestellar gas distributions in Figures \label{fig:gas2} and \label{fig:gas3} are more extended and elongated. The clumpy nature of these distributions highlights how the accretion histories of high-mass stars can vary significantly over time. Note that the characteristic timescales of the clumpiness within the accreting material, $t \sim \ell / \sigma_{\rm rms} \sim 0.1-1$ Myr, are significantly different than the characteristic variability time-scales imposed by disk instability, where bursts may last $\sim 10^2$ years with duty cycles of $10^3$ years \citep[e.g.,][]{OffnerMcKee2011a,KratterLodato2016a}.

In a couple of the panels the gas morphology appears to trace cloud-scale filamentary structures, e.g., the distribution of the high-mass accreting material in Fig.~\ref{fig:gas2}. Meanwhile, the filamentary structures on smaller scales shown in Fig.~\ref{fig:gas1} are reminiscent of ``streamers" \citep{PinedaArzoumanian2023a}, elongated structures that funnel material from the parent core down to the circumstellar accretion disk.

\begin{figure}
    \centering
    \includegraphics[width=0.99\linewidth]{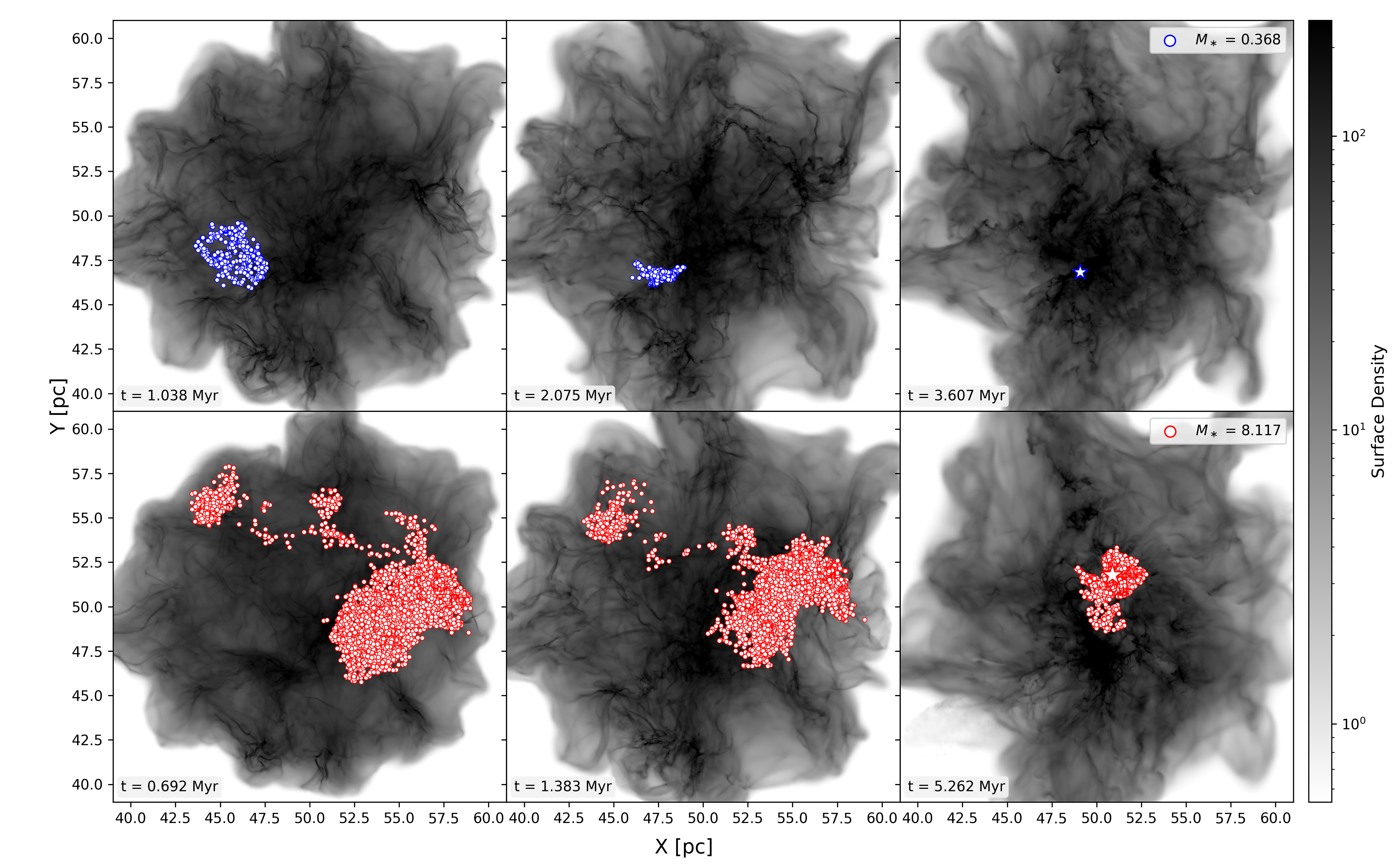}
     \includegraphics[width=0.99\linewidth]{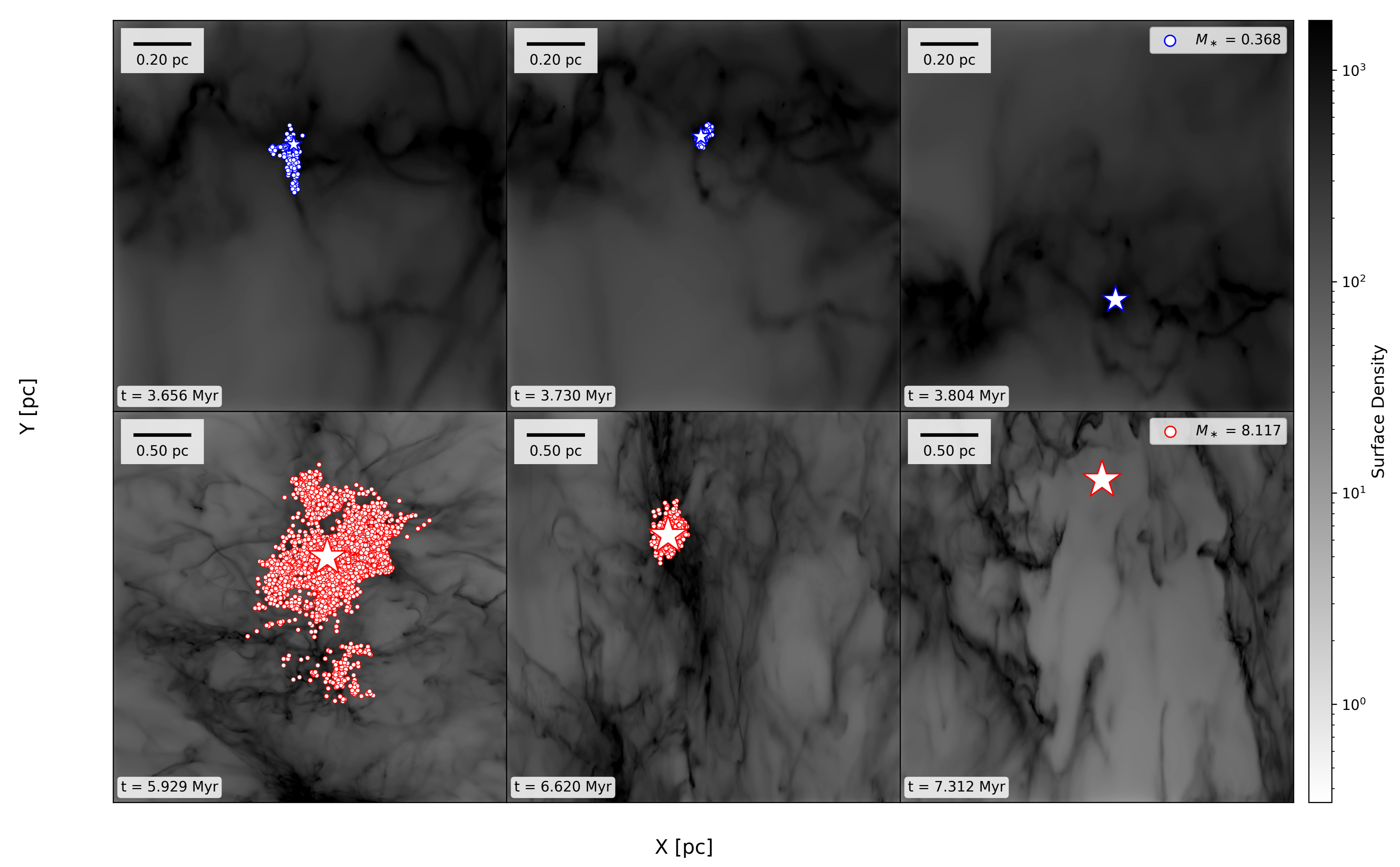}
    \caption{   Top panels: Same as Figure \ref{fig:ex_reservoir_pre} but for stars with final masses of $0.368\text{ }M_{\odot}$ (top) and  $8.117\text{ }M_{\odot}$ (bottom).  Bottom panels: Same as Figure \ref{fig:ex_reservoir_post} but for the two stars shown in the top panels.  The stars are forming in run \med. \label{fig:gas2}}
\end{figure}


\end{document}